\pdfoutput=1

\documentclass[aps,twocolumn,amsmath,amssymb,preprintnumbers,floatfix,prl,superscriptaddress,longbibliography]{revtex4-1}

\usepackage[utf8]{inputenc}
\usepackage{newtxtext}
\usepackage[upint]{newtxmath}
\usepackage{microtype}
\usepackage{textcomp}
\usepackage{eucal}
\usepackage{bm}
\usepackage{siunitx}
\usepackage{comment}
\usepackage{lipsum}

\usepackage{enumerate}
\usepackage{amsfonts}
\usepackage{amsmath}
\usepackage{amssymb}
\usepackage{color}
\usepackage{soul}

\usepackage{graphicx}

\usepackage[colorlinks,allcolors=blue]{hyperref}
\usepackage[capitalize]{cleveref}

\definecolor{DarkRed}{rgb}{0.65,0,0}%
\definecolor{Green}{rgb}{0,0.3,0.3}
\definecolor{Purple}{rgb}{0.3,0,0.65}
\definecolor{Red}{rgb}{1,0,0}
\definecolor{Blue}{rgb}{0,0,0.85}
\definecolor{Magenta}{rgb}{1,0,1}

\newcommand{\etal}{\emph{et al.}}

\newcommand{\be}{\begin{equation}}
\newcommand{\ee}{\end{equation}}

\newcommand{\prlsection}[1]{\textit{#1}.\kern0.05em---\kern0.05em\ignorespaces}

\begin{document}
\title{On-off switch and sign change for non-local Josephson diode in spin-valve Andreev molecules}
\author{Erik Wegner Hodt}
\email[Corresponding author: ]{erik.w.hodt@ntnu.no}
\affiliation{Center for Quantum Spintronics, Department of Physics, Norwegian \\ University of Science and Technology, NO-7491 Trondheim, Norway}
\author{Jacob Linder}
\affiliation{Center for Quantum Spintronics, Department of Physics, Norwegian \\ University of Science and Technology, NO-7491 Trondheim, Norway}

\begin{abstract}
Andreev molecules consist of two coherently coupled Josephson junctions and permit non-local control over supercurrents. By making the barriers magnetic and thus creating a spin-valve, we predict that a non-local Josephson diode effect occurs that is switchable via the magnetic configuration of the barriers. The diode effect is turned on, off, or changes its sign depending on whether the spin-valve is in a parallel, normal, or antiparallel configuration. These results offer a way to exert complete control over a non-local Josephson diode effect via the spin degree of freedom rather than varying a global magnetic flux which affects the entire system and likely neighbouring components in a device architecture. 
\end{abstract}
\maketitle

The flow of a supercurrent between superconductors separated by a non-superconducting restriction, the Josephson effect \cite{Likharev}, is a striking depiction of the quantum nature of the superconducting state and its physical implementation, the Josephson junction, is a fundamental device in quantum technology applications such as magnetic field sensing and metrology \cite{Wendin2017, Fagaly2006, Jeanneret2009181}. 

The interaction between multiple Josephson junctions (JJs) located within a distance on the order of the superconducting coherence length $\xi_0$ is an emerging field of interest. While there are numerous works on diverse non-local effects in systems with several Josephson junctions, several works have considered a particular model system known as the Andreev molecule, both theoretically \cite{nanoletters, DrivenAndreevmolecule, fineenergysplitting, scatteringdesc, Zsurka_2023, Chamoli:2022vh, pillet2023josephson} and experimentally \cite{Matsuo:2022tc, matsuo2023phasedependent, Krtssy2021, Scherübl2019, andreev_molecule}. 
The Andreev molecule is formed by the hybridization of overlapping Andreev bound states (ABS) stemming from individual JJs separated by a distance on the order of $\xi_0$. The Andreev molecule has been predicted to depict a non-local Josephson effect due to the non-local interaction between the phase gradients over the two JJs \cite{nanoletters}, causing a deviation from the single-junction current-phase relation and a non-reciprocal critical current. This can be viewed as a non-locally induced superconducting diode effect.
\begin{figure}[htb!]
    \centering
    \includegraphics{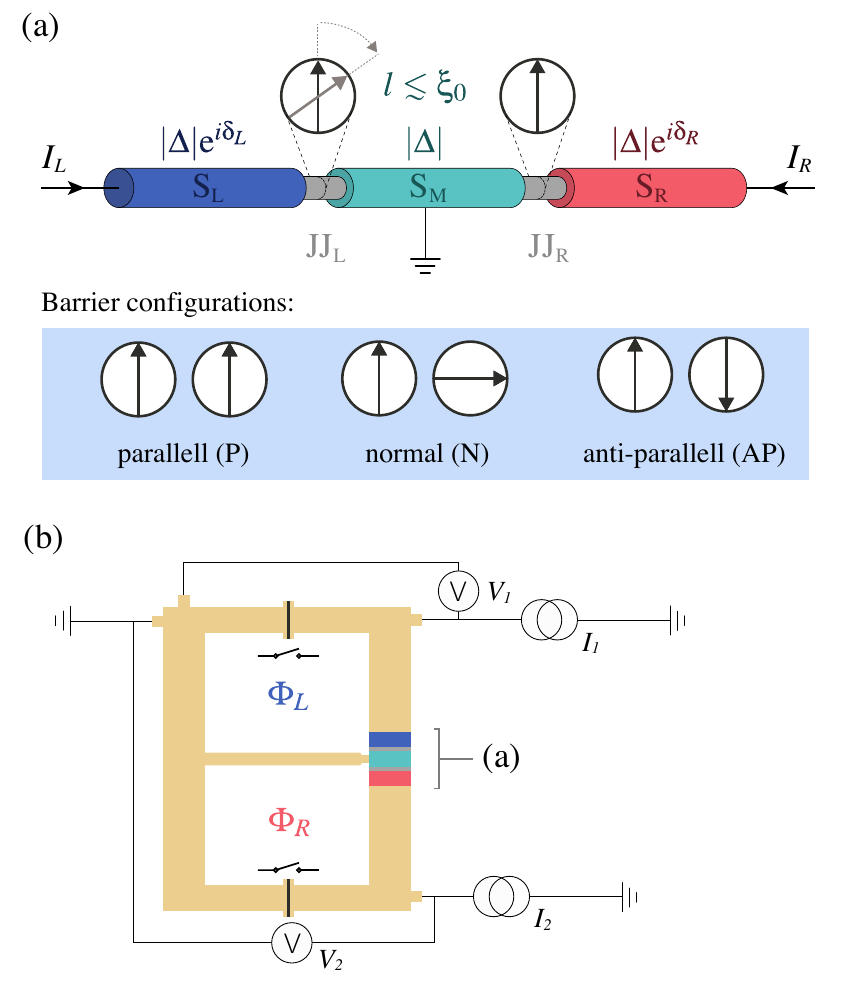}
    \caption{(a) The spin-valve Andreev molecule consists of three 1D superconductors where the phase difference between the superconducting order parameters is fixed. By a relative rotation of the magnetic moments of the spin-active barriers separating the three superconducting regions, drastically different superconducting diode characteristics can be achieved in the molecule.  (b) Experiment proposal for characterizing the spin-valve Andreev molecule. By pinching off the upper-most current loop, the phase difference between the middle and right superconductors can be fixed with a magnetic flux $\Phi_R$. }
    \label{fig: system configuration}
\end{figure}

The superconducting diode effect, the observation of an asymmetry between the forward and reverse critical currents $I_{c+}\neq I_{c-}$, is believed to be an important building block in future, dissipationless electronics devices. The effect was observed for a bulk system by Ando \etal \cite{nature_diodeeffect} and was attributed to a magnetochiral anisotropy, caused by the breaking of time and spatial inversion symmetries. Apart from properties of bulk superconductors, there have been several  predictions of rectifying behaviour in Josephson junction-based systems, for instance Refs. \cite{Reynoso, Zazunov, Yokoyama, Zazunov,Silaev2014, Dolcini, Chen, Minutillo, Pal2019, Kopasov} where the sources of the asymmetry are diverse, originating in magnetic barriers, sample geometry and the presence of spin-orbit coupling among others. As indicated by Pilet \etal \cite{nanoletters}, the non-local modulation of the current-phase relation in an Andreev molecule also introduces an asymmetry between the positive and negative critical current. As such, the Andreev molecule is an interesting system for the study and realization of the superconducting diode effect. However, the role of the spin degree of freedom has not been addressed so far in the literature on this system. 

In this Letter, we introduce the spin-valve Andreev molecule, consisting of three superconducting regions separated by spin-active barriers (see Fig. \ref{fig: system configuration}(a)). The relative orientation of the magnetic moment on the two barriers can be rotated in experimental setups and we show that the rectifying behaviour of the spin-valve Andreev molecule changes significantly when the magnetic barriers are parallel (P), normal to each other (N) or antiparallel (AP). We show that by switching the relative magnetization from P to (1) N  and (2) AP, the diode effect in the molecule can be (1) switched off and (2) reversed, with comparable diode efficiency in the other direction. This shows that the spin degree of freedom in coherently coupled Josephson junctions can be used to obtain new functionality, offering full control over the diode effect. 

\prlsection{Model}
The spin-valve Andreev molecule consists of three 1D superconductors connected by ferromagnetic weak links, as depicted in Fig. \ref{fig: system configuration}(a). The phase differences between the superconducting order parameters across each weak link are fixed, for the middle superconductor by a connection to ground. If the weak links are separated by a distance on the order of the superconducting coherence length $\xi_0$, the phase difference across one weak link can affect the ``effective" phase over the other through hybridization of the Andreev bound states (ABS) centred at each weak link. The spin-active barriers serving as the weak links are modelled as Dirac $\delta$-functions.

We model the SFSFS spin-valve Andreev molecule using the Bogoliubov-de Gennes formalism \cite{DeGennes2018}. Due to the spin-splitting induced by the ferromagnetic barriers, we consider the full $4\times 4 $ Nambu space Hamiltonian 
\begin{equation}
    H=\begin{pmatrix}
        H_0+V_{\uparrow\uparrow} & V_{\uparrow\downarrow} & 0 &  \Delta \\
        V_{\downarrow\uparrow} & H_0+V_{\downarrow\downarrow} & \Delta & 0 \\
        0 & \Delta^* & -H_0-V_{\uparrow\uparrow} & -V_{\uparrow\downarrow}^* \\
        \Delta^* & 0 & -V_{\downarrow\uparrow}^* & -H_0-V_{\downarrow\downarrow}
    \end{pmatrix}
    \label{eqn: Hamiltonian}
\end{equation}
where the gap parameter $\Delta(x)$ and potential $V(x)$ are defined by 
\begin{gather}
    \Delta(x)=\begin{cases}
        |\Delta |{e}^{i\delta_L} & \text{if } x < -l/2 \\
        0 & \text{if } |x| < l/2 \\
        |\Delta|{e}^{i\delta_R} & \text{if } x > l/2
    \end{cases} \\
    \begin{split}
    V =& U_L(\sigma_0+\gamma\hat{n}_L\cdot\boldsymbol{\sigma})\delta(x+l/2) \\ &\qquad +U_R(\sigma_0+\gamma\hat{n}_R\cdot\boldsymbol{\sigma})\delta(x-l/2)
    \end{split}
\end{gather}
where $H_0=\frac{-\hbar^2}{2m}\partial_x^2-\mu$, $l$ is the length of the middle superconductor, $U_0$ is the spin-independent barrier potential and $0.0 < \gamma < 1.0$ denotes the strength of the spin-active potential relative to $U_L$ / $U_R$. Moreover, $m$ is the electron mass, $\mu$ the chemical potential, $\hat{n}_{L/R}$ the unit vector denoting the direction of the left/right barrier moment, and $\boldsymbol{\sigma}$ the vector of Pauli matrices while $\sigma_0$ is the identity matrix. We will only consider situations where the spin-independent barriers are symmetric, $U_L = U_R = U_0=0.25\hbar v_F$ where $v_F$ is the Fermi velocity. For the non-magnetic Andreev molecule, this barrier strength corresponds to a transmission probability of $\tau=0.94$ which is considered realistic for single-channel conductors such as InAs-Al nanowires \cite{Goffman_2017}.  

The diode efficiency is quantified by the difference in critical supercurrent in the positive and negative direction, $\Delta I_c= I_{c+}-I_{c-}$ and the diode efficiency is commonly defined as \cite{unijosephsondiode}
\begin{equation}
    \eta \equiv \frac{\Delta I_c}{I_{c+}+I_{c-}}
\end{equation}

Solving Eq. (\ref{eqn: Hamiltonian}) in each of the three superconductors and using appropriate boundary conditions at the interfaces (see supplemental information for details), one obtains the discrete sub-gap ($E<|\Delta |$) energy levels known as Andreev bound states as well as a continuum of states for energies $E>|\Delta|$. We make the common semi-classical approximation $\xi_0 \gg \lambda_F$ where $\lambda_F$ is the Fermi wavelength and choose the Fermi momentum such that $k_F l$=$\pi/2 \text{ mod } 2\pi$, $k_F l \gg 1$. The two barriers of the Andreev molecule form an effective Fabry-Perrot resonator and the transmission $\tau$ is affected by whether $k_F l = 0 \text{ (mod }2\pi)$ (on-resonance) or $k_F l = \pi/2 \text{ (mod }2\pi)$ (off-resonance) \cite{nanoletters,scatteringdesc}. The results below are obtained in the off-resonance condition (the difference between on- and off-resonance is discussed in the supplemental information). For the spin-valve Andreev molecule, the coupling of the independent plane wave solutions in each superconductor involves solving for  16 coefficients giving the weight of the electron and hole-type plane waves of spin-up / spin-down in the three superconductors. The results shown in this paper are calculated numerically.

\begin{figure}[htb]
    \centering
    \includegraphics{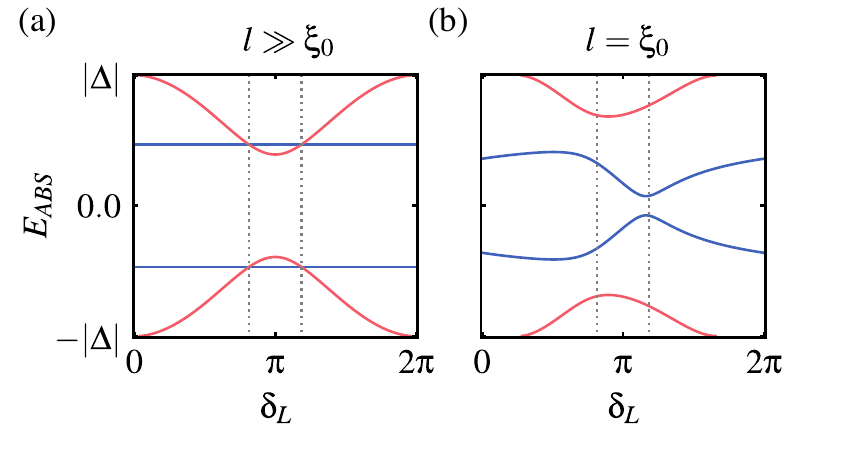}
    \caption{Andreev bound states (ABS) for the Andreev molecule with the length of the middle superconductor (a) significantly longer than, or (b) comparable to the superconducting coherence length $\xi_0$. For $l\gg\xi_0$, $E_{ABS}(\delta_L, \delta_R)\rightarrow E_{ABS}(\delta_L)$ for the ABS localized at the left weak link as the non-local impact of $\delta_R$ diminishes. As $l$ becomes on the order of $\xi_0$, the ABS from the two junctions hybridize into Andreev molecule orbitals. These states feature avoided crossings at $\delta_L=\pm\delta_R$ (dotted lines) related to the emergence of double-crossed Andreev reflection (dCAR) and double elastic co-tunnelling (dEC) processes. 
    }
    \label{fig: non-locality}
\end{figure}

\prlsection{Results and discussion}
\begin{figure*}[t]
    \centering
    \includegraphics{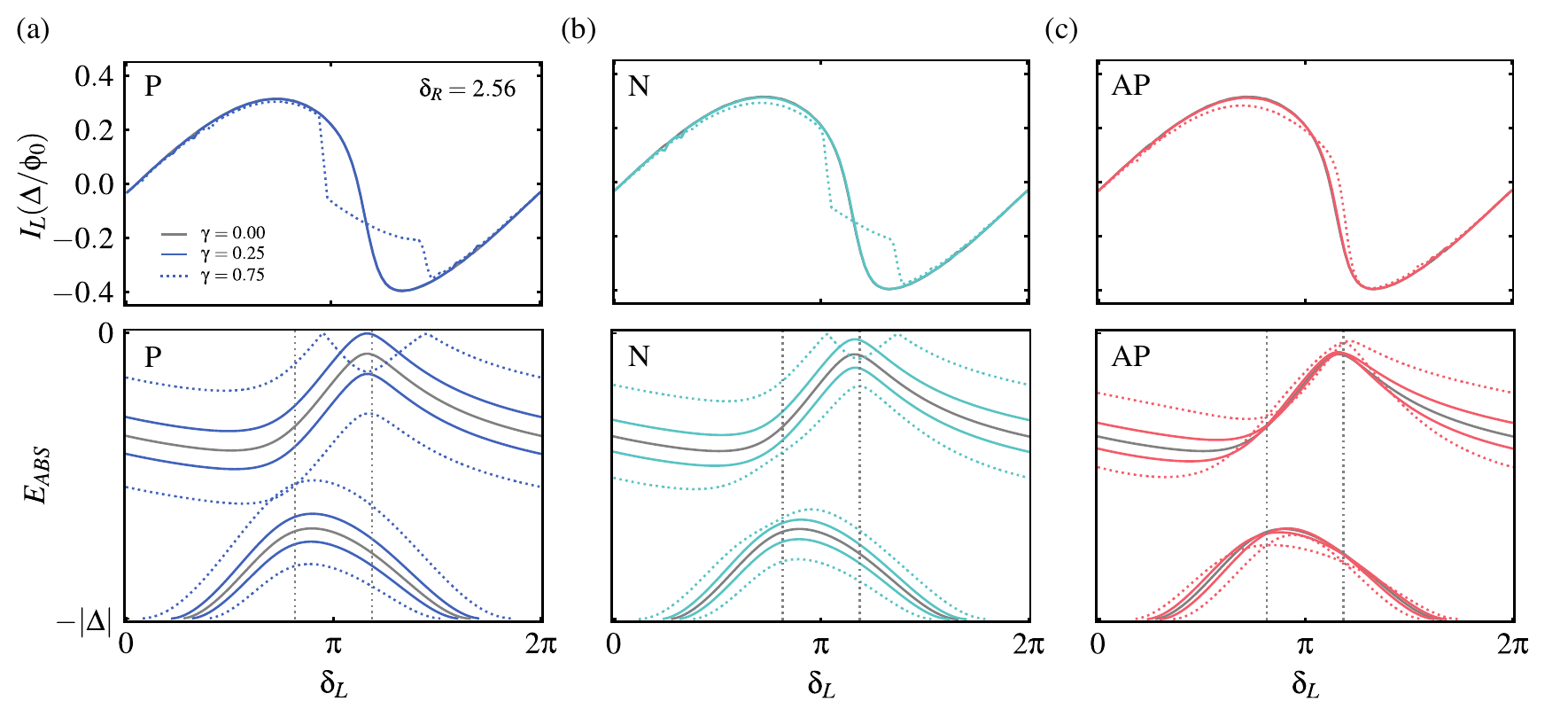}
    \caption{Current through the left junction as well as the sub-gap Andreev bound states (ABS) as a function of local phase $\delta_L$ for barrier strengths ($\gamma=0.00)$, $\gamma=0.25$ and $\gamma=0.75$ for the parallel (a), normal (b) and antiparallel (c) configuration. The non-local phase is set at $\delta_R=2.56$. As the strength of the magnetic barrier increases, the P ABS experiences a Zeeman-like spin-splitting, leading to current discontinuities caused by band-inversion as the gap between the ABS at the Fermi level closes at a critical value $\gamma \simeq 0.3$. In contrast, the AP configuration retains its spin-degeneracy at the points $\delta_L = \pm\delta_R$ (black dotted lines) and the Fermi level gap remains open for all parameters investigated but with a reduced $I_{c+}$ due to spin-splitting induced mismatch in the phase gradient of the four bands. The N configuration ABS depicts both the reduction in $I_{c+}$ from AP and the band inversion and current discontinuities from N, behaving as an effective mix of the two extrema.}
    \label{fig: ABS_SPREAD}
\end{figure*}
As the length of the middle superconductor in the Andreev molecule becomes on the order of the superconducting coherence length $\xi_0$, the ABS located at each of the two Josephson junctions hybridize.
The ABS for the regular Andreev molecule without magnetic barriers is shown in Fig. \ref{fig: non-locality} for middle superconductor lengths $l\gg\xi_0$ and $l=\xi_0$. For long separations ($l\gg\xi_0$), the molecule behaves as two independent junctions with a regular, short weak-link $I_c\propto\sin(\delta_L)$ current-phase relation for the left junction while the ABS from the separated, right junction remain dispersion-less under the variation of the local phase $\delta_L$. This reflects the absence of non-local modulation for this setup. As $l$ becomes on the order of a coherence length ($l=\xi_0$), the bands hybridize and avoided crossings arise at the previous band degeneracies at $\delta_L=\pm\delta_R$. These are caused by two distinct processes in the Andreev molecule, double-crossed Andreev reflection (dCAR) and double elastic co-tunnelling (dEC).  dEC facilitates transmission of Cooper pairs through the molecule and occurs for $\delta_L=-\delta_R$ (i.e. the same phase gradient over both junctions) \cite{Freyn} while dCAR involves the creation of a Cooper pair in the middle superconductor by an incident electron from the left in the left superconductor and an Andreev reflected hole in the right superconductor for $\delta_L=\delta_R$, made possible by the length of the middle superconductor being on the order of the extent of the Cooper pair itself \cite{Deutscher:2002uj, scatteringdesc}. 

The magnetic barriers of the spin-valve Andreev molecule lift the spin-degeneracy of the spin-up and spin-down states and have a significant effect on the ABS spectrum. The left junction current $I_L$, as well as ABS, is shown in Fig. \ref{fig: ABS_SPREAD} for the parallel (P), normal (N), and antiparallel (AP) configuration, as depicted in Fig. \ref{fig: system configuration}. The non-local phase is fixed at $\delta_R=2.56$, the value which gives the largest superconducting diode effect in the non-magnetic Andreev molecule. As the barriers become spin-active, the ABS in the P configuration are spin split in a manner reminiscent of regular Zeeman-type splitting where the band curvature remains largely unchanged, except close to the gap $-\Delta$. This is significant because the current contribution from the ABS spectrum is proportional to the phase gradient of the ABS bands. The AP configuration experiences a similar splitting away from the avoided crossings at $\delta_L=\pm\delta_R$, but due to a retained spin-degeneracy at those points, the phase gradient of the ABS is more affected in this case. An additional important consequence of the magnetic barriers is that for the P configuration, the gap at the Fermi level ($E_{ABS}=0.0$) closes for a critical barrier strength $\gamma\simeq0.3$. As a consequence, the ABS spectrum for the P configuration displays band inversion above this threshold, causing a cancellation of the current contribution due to the two upper bands in the ABS spectrum. For a regular Josephson junction, ABS spin splitting has no rectifying effect on the current-phase relation due to a symmetry around $\delta_L=\pi$. In the Andreev molecule, this symmetry is broken due to the non-local modulation from $\delta_R$, causing the band inversion to have an impact also on the rectifying behaviour of the non-local Josephson effect. 

\begin{figure*}
    \centering
    \includegraphics{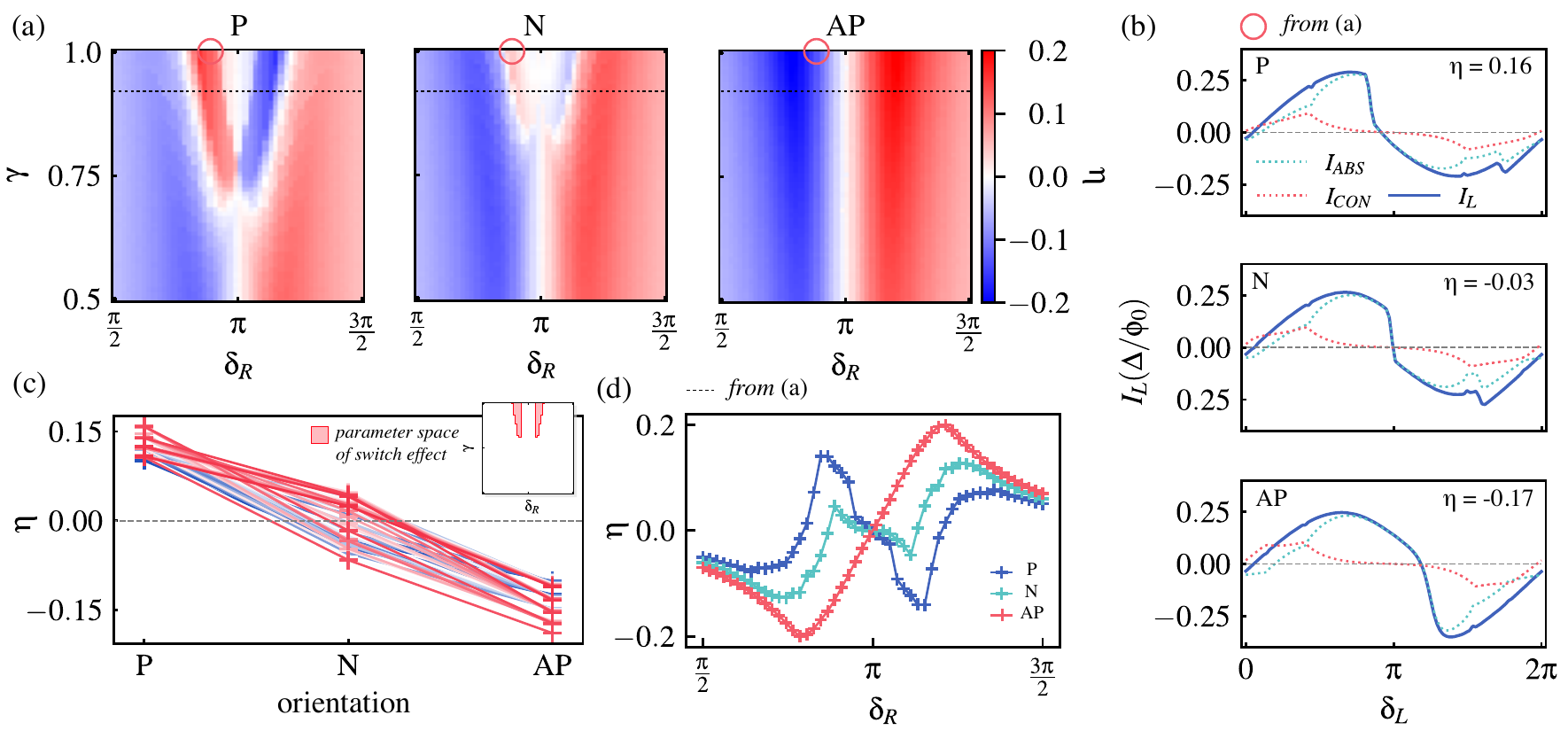}
    \caption{On-off switch as well as sign change in the non-local Josephson diode efficiency is shown for a spin-valve Andreev molecule with $l=\text{ }\xi_0$. (a) The diode efficiency dependence on the non-local phase $\delta_R$ and spin-splitting of the barrier $\gamma$ shows that significant deviations arise between the three configurations for $\gamma>0.5$. (b) The JJ$_\text{L}$ current-phase relation is shown for the parameter region denoted by red rings in (a). For specific combinations of $\delta_R$ and $\gamma$ shown in (c), $\eta > 0.1$ for the P configuration, $-0.1 < \eta < 0.1$ for N and $\eta < -0.1$ for the AP configuration. This enables an effective on-off switching as well as sign change of the diode efficiency by an appropriate rotation of the spin-valve configuration.  (d) The diode efficiency is shown as a function of non-local phase $\delta_R$ along the dotted line in (a), for $\gamma\simeq0.9$, highlighting the asymmetry between the three configurations.}
    \label{fig: diode_results}
\end{figure*}

We now demonstrate that the combination of the magnetic barriers and the non-local current-phase modulation in the spin-valve Andreev molecule constitutes a novel route for exerting complete control of the superconducting diode effect: both its existence and its sign can be tuned \textit{in situ} via the spin-valve configuration. The diode efficiency in the critical current through the left junction is shown as a function of the non-local phase $\delta_R$ in Fig. \ref{fig: diode_results} for a system with length $l=\xi_0$. For values of the spin-active barrier strength $\gamma<0.5$ (see additional data in supplemental information), the three barrier configurations show similar diode efficiency characteristics, but a significant asymmetry in the diode behaviour arises above $\gamma=0.5$. While the diode efficiency of the AP configuration increases monotonously with $\gamma$, the diode efficiency of the P configuration depicts a reversion phenomenon for $\delta_R$ close to $\pi$, as is evident from Fig. \ref{fig: diode_results}(a). The combined effect of the breaking of symmetry in the current-phase relation around $\delta_L=\pi$ as well as the barrier-induced band inversion and subsequent partial cancellation of the ABS phase gradients for $\delta_L>\pi$ cause the negative critical current $I_{c-}$ to reduce in magnitude for the P configuration. This makes the originally negative diode efficiency zero and then positive for $\gamma>0.5$ in the interval $\frac{3\pi}{4} < \delta_R < \pi$. The behaviour for $\delta_R>\pi$ is equivalent but with the opposite sign. For the N configuration, the combination of a weaker band inversion effect as well as a reduction in critical positive current $I_{c+}$ (see Fig. \ref{fig: ABS_SPREAD}(b) upper panel) effectively establishes a region of vanishing diode efficiency. The junction current in this parameter region is shown for the P, N and AP configuration in Fig. \ref{fig: diode_results}(b). 

As a metric of the extent of this effect, we show the region of $\delta_R$ and $\gamma$ for which an on-off effect as well as switching with a P/AP threshold efficiency of $\eta_\text{lim}>|0.1|$ and N efficiency of $-0.1<\eta_{\text{lim}}<0.1$ is achievable. Fig \ref{fig: diode_results} (c) shows the diode efficiency for these specific $\delta_R$\ and $\gamma$ and their respective efficiency in the parallel, normal and anti-parallel configuration. The inset shows where in the diagrams in Fig. \ref{fig: diode_results} (a) both the on-off and switching effects are observed. We note that this parameter region increases significantly if the threshold efficiency $\eta_\text{lim}$ is lowered or if one relinquishes the need for a vanishing normal diode efficiency and only considers the switching between positive and negative diode efficiency for the P and AP configuration. 

To conclude, we propose an on-off and switching mechanism in the superconducting diode effect of a spin-valve Andreev molecule. This occurs due to the interplay between non-local phase modulation from the regular Andreev molecule and the introduction of magnetic barriers which alters the ABS spectrum of the molecule. The combination of a vanishing Fermi level gap and subsequent band inversion with non-local phase modulation in the parallel configuration Andreev molecule causes a reversion of the diode efficiency which establishes a parameter region where the magnitude and sign of the diode efficiency can be tuned by relative rotation of the magnetic barrier moments. This entails the possibility of a device where one can switch on-off as well as reverse the diode efficiency for a constant phase-bias $\delta_R$. 

 \begin{acknowledgments}
We thank C. Br{\"u}ne and N. Birge for helpful comments. This work was supported by the Research
Council of Norway through Grant No. 323766 and its Centres
of Excellence funding scheme Grant No. 262633 “QuSpin.” Support from
Sigma2 - the National Infrastructure for High Performance
Computing and Data Storage in Norway, project NN9577K, is acknowledged. The research presented in this paper has benefited from the Experimental Infrastructure for Exploration of Exascale Computing (eX3), which is financially supported by the Research Council of Norway under contract 270053.
 \end{acknowledgments}

\bibliography{refs.bib}

\appendix

\end{document}